\documentclass[aps,prd,twocolumn,superscriptaddress]{revtex4-1}
\usepackage[]{mathtools}
\usepackage{bm}
\usepackage{graphicx}
\usepackage{subfigure} 
\usepackage{float}
\usepackage[colorlinks,linkcolor=blue,anchorcolor=blue,citecolor=blue]{hyperref} % hyper link
\usepackage{verbatim}
\usepackage{booktabs}
\usepackage{multirow}
\usepackage{color}
\usepackage{mathtools}
\usepackage{longtable}
\usepackage{comment}
\usepackage{amssymb}
\usepackage{bm}
\begin{document}

%\preprint{}

\title{Costs of Bayesian Parameter Estimation in Third-Generation Gravitational Wave Detectors: an Assessment of Current Acceleration Methods}

\author{Qian Hu}
%\email{q.hu.2@research.gla.ac.uk}
\email{Qian.Hu@glasgow.ac.uk}
\affiliation{Institute for Gravitational Research, School of Physics and Astronomy, University of Glasgow, Glasgow, G12 8QQ, United Kingdom}

\author{John Veitch}
\email{John.Veitch@glasgow.ac.uk}
\affiliation{Institute for Gravitational Research, School of Physics and Astronomy, University of Glasgow, Glasgow, G12 8QQ, United Kingdom}

\date{\today}

\begin{abstract}
    Bayesian inference with stochastic sampling has been widely used to obtain the properties of gravitational wave (GW) sources. Although computationally intensive, its cost remains manageable for current second-generation GW detectors because of the relatively low event rate and signal-to-noise ratio (SNR). The third-generation (3G) GW detectors are expected to detect hundreds of thousands of compact binary coalescence (CBC) events every year with substantially higher SNR and longer signal duration, presenting significant computational challenges. In this study, we systematically evaluate the computational costs of {CBC} source parameter estimation (PE) in the 3G era by modeling the PE time cost as a function of SNR and signal duration. We examine the standard PE method alongside acceleration methods including relative binning, multibanding, and reduced order quadrature. We predict that PE for a one-month-observation catalog with 3G detectors could require {at least billions} of CPU core hours with the standard PE method, whereas acceleration techniques can reduce this demand to {less than millions} of core hours, {which is as high as the cost of analyzing GW events in the past 10 years}. These findings highlight the necessity for more efficient PE methods to enable cost-effective and environmentally sustainable data analysis for 3G detectors. In addition, we assess the accuracy of accelerated PE methods, emphasizing the need for careful treatment in high-SNR scenarios.
\end{abstract}

\maketitle
\section{\label{sec1}Introduction}
% LIGOScientific:2018mvr, LIGOScientific:2020ibl,  
Since the first direct detection of gravitational waves (GWs) by Advanced LIGO~\cite{LIGOScientific:2014pky} in 2015~\cite{LIGOScientific:2016aoc}, {over 200} GW events have been published by the LIGO-Virgo\cite{VIRGO:2014yos}-KAGRA\cite{KAGRA:2020tym} (LVK) collaboration~\cite{LIGOScientific:2025slb, LIGOScientific:2021usb, KAGRA:2021vkt}, with hundreds more expected from the current observing run. All of these GW events are identified as compact binary coalescences (CBCs) - mergers of compact objects such as black holes, neutron stars, and potentially exotic stars~\cite{Liebling:2012fv, Cardoso:2019rvt}. Determining the properties of these sources is essential for various aspects of GW astronomy, including cosmology~\cite{LIGOScientific:2017adf, DES:2019ccw, LIGOScientific:2019zcs}, astrophysical population~\cite{LIGOScientific:2018jsj, LIGOScientific:2020kqk, KAGRA:2021duu}, the equation of state of dense matter~\cite{LIGOScientific:2018cki}, stochastic background~\cite{LIGOScientific:2016jlg, LIGOScientific:2016nwa, LIGOScientific:2017zlf}, among others.

CBC signals are characterized by the physical and geometric parameters of their sources, and the properties of the sources are inferred by estimating these parameters.
A typical CBC system is described by at least 15 parameters, including two component masses, six component spin parameters, and seven extrinsic parameters describing the 3D sky location, inclination angle and polarization angle, and coalescence phase and time of the source. Additional parameters are required to account for specific phenomena, such as tidal deformability parameters for neutron stars~\cite{PhysRevD.89.103012}, orbital eccentricity and eccentric anomaly for non-quasi-circular orbits~\cite{Shaikh:2023ypz, Ramos-Buades:2023yhy}, and modification terms for alternative theories of gravity~\cite{Mishra:2010tp, LIGOScientific:2021sio}, etc. 

Due to the transient nature of CBC signals, parameter estimation (PE) is solved under the Bayesian framework~\cite{Veitch:2014wba}, in which the probability density distribution of the source parameters is given by the Bayes theorem, but not directly analyzable {due to the complexity in GW waveform models. }
Stochastic sampling algorithms, Markov Chain Monte Carlo (MCMC) or Nested Sampling, are used to draw samples from the probability distribution, which represent the estimated source parameters.
These methods, however, are computationally intensive. Currently, a typical {full PE run using the standard Bayesian framework~\cite{Veitch:2014wba} for a CBC source} would cost hours to days, and the cost increases with the signal-to-noise ratio (SNR) and signal duration. A number of acceleration methods for GW source PE are proposed to mitigate the computational burden in stochastic sampling, including Relative Binning (RB, also referred to as Heterodyning)~\cite{Zackay:2018qdy, Leslie:2021ssu, Cornish:2010kf}, Multibanding (MB)\cite{Vinciguerra:2017ngf, Morisaki:2021ngj}, and Reduced Order Quadrature (ROQ)~\cite{Canizares:2014fya, Smith:2016qas}. Nevertheless, little work has been done to systematically investigate and compare the performance of these algorithms. The LVK data release still extensively uses the standard PE method without acceleration~\cite{LIGOScientific:2018mvr, LIGOScientific:2020ibl, LIGOScientific:2021usb, KAGRA:2021vkt}, {which is feasible given} current detection rate, signal SNRs and durations. 
 
However, there is a growing concern regarding the computational burden for the proposed third-generation (3G) GW detectors~\cite{Couvares:2021ajn, Bagnasco:2023ren}. The 3G detectors, including Einstein Telescope (ET)~\cite{Punturo:2010zz, Branchesi:2023mws, Abac:2025saz} and Cosmic Explorer (CE)~\cite{reitze2019_CosmicExplorerContribution, Evans:2023euw}, are designed to %reach the theoretical limit of sensitivity of ground-based GW detectors
{improve the sensitivity by at least an order of magnitude over Advanced LIGO},
and are expected to detect hundreds of thousands of CBC signals every year with significantly higher SNR and longer signal duration~\cite{Branchesi:2023mws, Borhanian:2022czq}. While these detectors promise exciting scientific discoveries, the computational cost of analyzing all the detected signals using current methods is anticipated to be enormous, though it is not yet fully quantified. 

In this paper, we present a detailed estimate of the computational costs of PE {for CBC sources} in the 3G era, considering both the standard PE method and accelerated methods. By conducting 1200 PE experiments and fitting the PE time cost as a function of SNR and duration, we estimate the total costs of PE for 3G detectors based on the simulated catalog in ET Mock Data Challenge 1 (MDC-1)~\cite{Tania:2025bsa}. We find that billions of CPU core hours is required to analyze one month of observation with the 3G detectors using the standard PE method, with most of the computational load coming from long binary neutron star (BNS) and neutron star-black hole (NSBH) signals.
With acceleration methods, the cost could be reduced to {less than} millions of CPU core hours. 
We compare the performance of different acceleration methods in terms of speed and accuracy. Our results identify ROQ as the fastest method overall, though it requires additional training before the PE. {We find that, even with the accelerated methods, the one-month cost is sufficient to analyze all CBC events detected in the past ten years, highlighting the necessity of the use of more efficient methods such as machine learning.}
We also highlight accuracy issues in high-SNR scenarios, with RB showing the worst accuracy due to its reliance on fiducial parameters used in the heterodyning process.

This paper is organized as follows. Section~\ref{sec2} provides a brief introduction to Bayesian inference and stochastic sampling in Sec.~\ref{sec2}, including the standard methods in Sec.~\ref{sec21} and acceleration methods in Sec.~\ref{sec22}. The experiments are presented in Sec.~\ref{sec3}, with details on the experiment design in Sec.~\ref{sec31}, fitting the PE cost in Sec.~\ref{sec32}, and accuracy assessment in Sec.~\ref{sec33}.
Section~\ref{sec4} applies the time cost fitting to the ET MDC-1 and estimates the total PE cost in the 3G era. The concluding remarks are provided in Sec.~\ref{sec5}. {Throughout this paper, the phrase ``GW sources'' refers to CBCs.}

\section{\label{sec2}Overview of Bayesian parameter estimation}
\subsection{\label{sec21} Bayesian parameter estimation}
\subsubsection{\label{sec211} Bayes theorem}
Bayesian inference gives us means to quantify uncertainty about the source parameters using the concept of Bayesian updating. {Here we give a brief overview of the Bayesian inference framework used in GW astronomy and in this work. More detailed reviews of this topic can be found in, e.g, Refs~\cite{Veitch:2014wba, 2019PASA, Christensen:2022bxb}} Starting with generic prior expectations about the source (which may be informed by astrophysical or geometrical considerations), we observe the transient GW signal, and then update our knowledge based on the new data. Mathematically, given a hypothesis $H$ (e.g. the waveform model) and a prior distribution of the parameters $p(\theta | H)$, %(which can be non-informative)
we can update the probability distribution of $\theta$ with observation data $d$ using Bayes' theorem:
\begin{equation}
    \label{eq:posterior}
    p(\theta | d, H) = \frac{p(\theta | H) p(d|\theta, H)}{p(d|H)},
\end{equation}
where 
\begin{equation}
    \label{eq:likelihood}
    p(d|\theta, H) \propto \exp{-\frac{1}{2}(d-h(\theta)| d-h(\theta))}
\end{equation}
is the likelihood function, assuming the noise is stationary and Gaussian~\cite{Finn:1992wt}, $h(\theta)$ is the GW signal given parameter $\theta$, {and $(\cdot|\cdot)$ denotes the inner product between two strain data sets. The raw GW data is provided in the time domain, but the inner product is easier to estimate in the frequency domain using the windowed and Fourier-transformed data~\cite{Finn:1992wt}:}
\begin{equation}
    (a|b) =  4\,\mathrm{Re}\int_{0}^{\infty} \frac{a^*(f)b(f)}{S_n(f)} df , 
\end{equation}
where $\mathrm{Re}$ denotes real part and ${a}^*$ denotes the complex conjugate of $a$. $S_n(f)$ is the {one-sided} power spectral density (PSD) of the detector noise. $p(d|H)$ is the evidence, a constant that is irrelevant for the shape of the posterior distribution, but is useful in model comparison~\cite{2019PASA}. The posterior probability distribution $p(\theta | d, H)$ provides the estimation of the parameter $\theta$. 

\subsubsection{\label{sec212} Stochastic sampling}
Although the posterior probability density function is computable in the entire parameter space (as long as the waveform model is valid) via Eq.~\ref{eq:posterior}, it is impossible to compute it all over the space due to the high dimensionality. Instead, stochastic sampling algorithms are used to obtain a set of samples that follows the distribution given by Eq.~\ref{eq:posterior}.

MCMC and nested sampling are two families of sampling algorithms. %MCMC has a set of walkers randomly walking in the parameter space, with a higher probability of walking towards regions of higher target probability. Over a sufficiently long period of iterations, the trace of the walkers converges to the target probability distribution, which is designed to be the posterior probability.
{MCMC constructs a Markov chain whose equilibrium distribution is the target distribution. Over time, the chain wanders through the parameter space, producing samples with a density that approximates the target distribution.}
{Nested sampling is designed to calculate the evidence integral, the denominator of Eq.~\ref{eq:posterior}~\cite{skilling2004}. It works by initially drawing a set of `live points' from the prior distribution, then iteratively removing and replacing the sample with the lowest likelihood. As the algorithm progresses the live points occupy a geometrically shrinking volume of parameter space, whose volume can be statistically estimated. After the algorithm reaches the peak, the nested samples can be reweighted to draw or approximate the posterior distribution.}
%Nested sampling works with a set of live points drawn from the prior distribution. The live point with the lowest likelihood value will be discarded and a new point will be drawn from the space expanded by the remaining live points. Repeating this process, the live points will eventually form the posterior space. 

A number of variants of MCMC and nested sampling algorithms have been used in GW astronomy. \texttt{dynesty}~\cite{Speagle_2020} is the most widely used in recent LVK public data~\cite{LIGOScientific:2021usb, KAGRA:2021vkt}, although many novel samplers are being developed to achieve faster speeds. For example, \texttt{nessai}~\cite{Williams:2023ppp, Williams:2021qyt} is a nested sampling algorithm enhanced with machine learning and importance sampling, and it reduces the number of likelihood evaluations by an order of magnitude in GW source PE compared with \texttt{dynesty}. {We find that \texttt{dynesty} { {v2.1.4}}~\cite{sergey_koposov_2024_12537467} with the configuration used in the LVK production in the third observing run performs extremely slowly in our 3G simulations, while \texttt{nessai} {{v0.8.1}}~\cite{michael_j_williams_2023_7828161}} offers significantly improved speed and convergence without any fine-tuning. Therefore, we use {\texttt{nessai}} throughout this work.
%and leave the 3G configuration of \texttt{dynesty} to future investigations.}

\subsection{\label{sec22} Acceleration methods}
Stochastic sampling for GW PE is computationally intensive. Sampling algorithms may struggle to identify the correct region in the parameter space, requiring long convergence times and numerous likelihood evaluations (waveform generation and vector manipulation). 
An increase in signal duration {(which results in more observed GW cycles and a boarder bandwidth)} or SNR results in a narrower posterior {(see, e.g., Ref.~\cite{Cutler:1994ys, Fairhurst:2009tc})}, making it more challenging for the sampler to locate and converge on the posterior. Additionally, longer signal durations slow down waveform generation and vector manipulation involved in the likelihood evaluation. Both of these factors increase the computational cost of performing PE.

Several acceleration methods have been proposed to speed up the PE. In addition to the general improvement of the sampler like \texttt{nessai}, there are methods specifically targeting the CBC PE problem, aiming to reduce the data size and simplify likelihood evaluations. In this section, we introduce three methods that we will examine in this work: Relative Binning (RB), Multibanding (MB), and Reduced Order Quadrature (ROQ). All of them have been implemented in the GW PE Python package \texttt{bilby}~\cite{Ashton:2018jfp, Krishna:2023bug, Morisaki:2021ngj, Smith:2016qas}. 
We will also mention other fast PE methods in Sec.~\ref{sec224} but they are not involved in our assessment due to either having completely different mechanisms, being a mixture of many methods, only obtaining a subset of parameters, or having an incomplete software implementation. 

\subsubsection{\label{sec221} Relative Binning}
Relative Binning (RB)~\cite{Zackay:2018qdy, Leslie:2021ssu}, { first proposed as} heterodyning~\cite{ Cornish:2010kf}, is based on the idea that many likelihood evaluations occur near the maximum likelihood point during PE, and that the GW waveform changes smoothly as a function of the source parameters. Therefore, if a fiducial source parameter that is close to the true value can be assigned before PE (which is possible given information from matched filtering detection, {see, e.g., Ref.~\cite{Villa-Ortega:2022qdo}}), the PE can focus on in the nearby region where likelihood evaluations can be simplified by exploiting the smooth variations in the waveform as the fiducial parameter is varied.
Since the smooth variation with respect to the fiducial waveform requires a lower bandwidth to describe,
%Additionally,
a coarser frequency resolution is sufficient to reconstruct waveforms near the fiducial waveform. 
%, as the differences between them are smooth.
This reduces the data size and therefore the computational cost of computing each likelihood. The new frequency bins with coarse frequency resolution can be chosen such that the GW signal only contains a few cycles within each bin, which is computable using the post-Newtonian (PN) phase expansion~\cite{Buonanno:2009zt}.

Considering frequency domain waveforms, we approximate the ratio between the fiducial waveform $h_0(f)$ and an arbitrary waveform $h(f)$ in a frequency bin b to the linear order
\begin{equation}
    r(f)=\frac{h(f)}{h_0(f)}=r_0(h, \mathrm{b})+r_1(h, \mathrm{b})\left(f-f_{\mathrm{m}}(\mathrm{b})\right)+\cdots, 
\end{equation}
where is $f_{\mathrm{m}}$ is the central frequency of the bin. $r_0$ and $r_1$ are bin-dependent coefficients that measure how $h$ deviates from $h_0$, and can be efficiently derived from the values of $r(f)$ at the edges of bins. Since $(d|d)$ is a constant, the likelihood (Eq.~\ref{eq:likelihood}) only requires the computation of $(d|h)$ and $(h|h)$, which can be approximated using $r(f)$. Define 
\begin{equation}
    \begin{aligned}
        A_0(\mathrm{b}) &= (d|h_0)_\mathrm{b}, \quad A_1(\mathrm{b}) = (d|(f-f_\mathrm{m})h_0)_\mathrm{b} \\
        B_0(\mathrm{b}) &= (h_0|h_0)_\mathrm{b},  \quad B_1(\mathrm{b}) = (h_0|(f-f_\mathrm{m})h_0)_\mathrm{b}, 
    \end{aligned}
\end{equation}
where $(\dots | \dots)_\mathrm{b}$ is the inner product inside the frequency bin b, i.e., the integral limits are the minimum and the maximum frequencies of the bin. $(d|h)$ and $(h|h)$ can be calculated as 
\begin{equation}
    \begin{aligned}
        (d|h)^{\mathrm{RB}} &= \sum_{\mathrm{b}} (d|h)_\mathrm{b} = \sum_{\mathrm{b}} (d|r h_0)_\mathrm{b} \\
        &\approx \sum_{\mathrm{b}}\left(A_0(\mathrm{b}) r_0^*(h, \mathrm{b})+A_1(\mathrm{b}) r_1^*(h, \mathrm{b})\right),
    \end{aligned}
\end{equation}
and similarly 
\begin{equation}
    \begin{aligned}
        (h|h)^{\mathrm{RB}} \approx \sum_{\mathrm{b}}\left(B_0(\mathrm{b})\left|r_0(h, \mathrm{b})\right|^2+2 B_1(\mathrm{b}) \mathrm{R e}\left[r_0(h, \mathrm{b}) r_1^*(h, \mathrm{b})\right]\right).
    \end{aligned}
\end{equation}
For any waveform $h$, it is only necessary to evaluate the waveform at the edges of the bins to obtain $r_0$ and $r_1$, which speeds up the likelihood evaluations. In addition, focusing on regions around the fiducial parameters significantly reduces the number of likelihood evaluations, although this may bring accuracy issues. 

Examples of applications of RB can be found in Refs~\cite{Cornish:2021wxy, Roulet:2024hwz, Dai:2018dca,Dax:2024mcn, Wong:2023lgb}. In particular, \citet{Dai:2018dca} showed that PE for GW170817 with aligned spin waveform model \texttt{IMRPhenomD\_NRTidal}~\cite{Dietrich:2019kaq} and \texttt{TaylorF2}~\cite{Buonanno:2009zt} and RB can be done with 150 CPU core hours. \citet{Wong:2023lgb}, also using aligned spin waveform model \texttt{IMRPhenomD (\_NRTidal)}, demonstrated a combination of RB and a gradient-based sampler, and achieved 2 hours of sampling time for GW150914 and 1 day for GW170817 with 400 CPU cores.

\subsubsection{\label{sec222} Multibanding}
The frequencies of signals from CBC sources increase during the binary inspiral. GW signals are present in the data from the low-frequency cutoff $f_\mathrm{min}$ of the GW detector ($\sim 20$\,Hz for current detectors~\cite{LIGOScientific:2014pky, VIRGO:2014yos, KAGRA:2020tym} and $\sim 5$\,Hz for 3G~\cite{Abac:2025saz, reitze2019_CosmicExplorerContribution}) up to the maximum frequency $f_\mathrm{max}$ of hundreds of Hz up to $\sim 2$\,kHz, depending on the source masses. The full-bandwidth GW data (i.e. strain data with sampling rate of 4096\,Hz or 16384\,Hz) is excessive for the low-frequency early inspiral stage, therefore the data can be adaptively down-sampled as long as sampling rate remains above the Nyquist frequency. This approach is known as multibanding or adaptive frequency resolution, {first proposed for source detection~\cite{Adams:2015ulm, Aubin:2020goo}.}

Several different MB schemes for PE have been proposed; see, e.g., Refs~\cite{Vinciguerra:2017ngf, Morisaki:2021ngj}. In this work, we adopt the scheme used in \citet{Morisaki:2021ngj}, in which the time-domain data is divided into bands with geometrically decreasing lengths $(T, T/2, T/4, \dots, 4s)$. The starting and ending frequencies of each band are chosen such that (i) the data in the time-domain band can cover the signal within the frequency band, with time-frequency relation given by the 0PN equation for the lowest mass in the prior and (ii) the waveform smoothly vanishes at the left edge of each time-domain band with the window function applied. The frequency bands and relevant coefficients are determined once the prior is set, requiring no additional calculations before or during PE.

MB leads to coarser frequency grids on which the GW waveform needs to be evaluated and thus reduces the data size, accelerating waveform generation and likelihood evaluation. This requires waveform models that can be calculated at any frequency, typically with closed-form expressions in the frequency domain, such as the \texttt{IMRPhenom} waveforms~\cite{Husa:2015iqa,Khan:2015jqa}. However, MB does not reduce the number of likelihood evaluations. \citet{Morisaki:2021ngj} shows that MB can speed up PE by a factor of 20-50 for signals starting from 20 Hz, and by a factor of 100--500 for signals starting from 5 Hz, in line with the asymptotic scaling $5f_\mathrm{max}/3f_\mathrm{min}$ found in \cite{Vinciguerra:2017ngf}. MB can be combined with ROQ~\cite{Morisaki:2021ngj, Morisaki:2023kuq} and RB~\cite{Dax:2024mcn, Hu:2024oen} to further enhance the performance. In addition to PE and detection, the similar idea has been explored in source localization~\cite{Hu:2023hos}.

\subsubsection{\label{sec223} Reduced Order Quadrature}
ROQ~\cite{Canizares:2014fya, Smith:2016qas} utilizes the reduced order modeling (ROM)~\cite{Field:2013cfa, Purrer:2014fza,  Blackman:2015pia} of GW waveforms, which aims to accelerate waveform evaluation. In the frequency domain, the ROM of a waveform model can be written as the linear addition of $N$ bases, where $N$ is much smaller than the original length of the waveform: 
\begin{equation}
    h(f ; \theta) \approx h^{\mathrm{ROM}}(f ; \theta)=\sum_{J=1}^N h\left(\mathcal{F}_J ; \theta\right) B_J(f) ,
\end{equation}
where $B_J(f)$ is a set of bases that span the signal space obtained via a greedy algorithm~\cite{Field:2013cfa} or singular value decomposition~\cite{Purrer:2014fza}. $\{ \mathcal{F}_J \}$ is a set of the most representative frequencies that can be used to reconstruct the entire waveform determined by empirical interpolation~\cite{Field:2013cfa}. The $h^{\mathrm{ROM}}(f ; \theta)$, as an approximation of the original waveform model $h(f;\theta)$, only requires waveform evaluation at $\{ \mathcal{F}_J \}$ and a set of predetermined bases $\{B_J(f)\}$, and is therefore faster to calculate, {at the cost of pre-computing the bases}. 
For low-dimensional parameter spaces, it is possible to build a surrogate model $h_J(\theta)$ to approximate the waveform value at a specific frequency node $h\left(\mathcal{F}_J ; \theta\right)$. $h_J(\theta)$ can takes the form of polynomials~\cite{Field:2013cfa}, splines~\cite{Purrer:2014fza}, or even neural networks~\cite{Thomas:2022rmc}. However, PE for precessing systems has a high-dimensional parameter space in which surrogate models cannot be easily built, hence it is not suitable for our task. We therefore still need a waveform model that can be evaluated at any frequencies, like we do in MB.

The ROQ likelihood can be expressed using the ROM. The non-constant terms in the likelihood are:
\begin{equation}
    (d|h)^{\mathrm{ROQ}} = \mathrm{Re} \sum_{J=1}^N h\left(F_J ; \theta\right) \omega_J\left(t_c\right),
\end{equation}
and 
\begin{equation}
    (h|h)^{\mathrm{ROQ}}  = \sum_{I=1}^N \sum_{J=1}^N h^*\left(F_I ; \theta\right) h\left(F_J ; \theta\right) \psi_{I J},
\end{equation}
where $\omega_J\left(t_c\right)$ and $\psi_{I J}$ are integration weights that depend on the basis functions, data, and PSD, and their definitions can be found in Ref.~\cite{Smith:2021bqc}. To perform ROQ, basis functions $\{B_J(f)\}$ need to be calculated under a prior distribution of the intrinsic parameters (which can be broader than the prior used in PE) and the integration weights need to be tailored to the specific data.
This precalculation stage can take anywhere from minutes to hours, depending on factors such as signal length and the prior range. 
However, the most computationally expensive step, basis construction, only needs to be done once for a wide range of sources, as long as the source falls within the prior used for basis construction. 
ROQ significantly accelerates the likelihood calculation by reducing it to linear and quadratic summations with waveforms evaluated at a reduced frequency grid. 

ROQ can bring speed-up factors of more than ten times depending on the signal properties, and is widely-used in GW PE, see e.g.~\cite{Smith:2021bqc, Canizares:2014fya, Morras:2023pug, Morisaki:2023kuq}.  Notably, {the speed up can goes up to thousands when the ROQ prior is fine-tuned using the information from detection~\cite{Morisaki:2020oqk}. Similarly,} Ref.~\cite{Smith:2021bqc} illustrates ROQ's application to long signals from BNS in 3G detectors: using a restricted ROQ prior, PE for a 90-minute-long BNS signal of SNR of 2400 can be done with 20 CPU hours of precalculation and 1600 CPU hours of sampling. Bayesian inference for such signals would be prohibitively slow ($>10^7$ CPU hours) using the standard PE method. 

\subsubsection{\label{sec224} Other methods}
There are a number of other approaches for fast PE. One common technique is marginalization, which reduces computational cost by integrating over certain parameters.
The luminosity distance, coalescence time, and coalescence phase (for the dominant $\ell=m=2$ mode of the signal) can be analytically marginalized in the posterior~\cite{Veitch:2014wba,2019PASA}, allowing them to be excluded from the sampling process and reconstructed after the other parameters are sampled. 
Taking the matched filtering SNR timeseries as input, numerical marginalization of more parameters is demonstrated in \texttt{cogwheel}~\cite{Roulet:2024hwz}, achieving minutes to a few hours of PE time on a single CPU. \texttt{RIFT}~\cite{lange2018rapidaccurateparameterinference, Wofford:2023iwz} also makes use of likelihood marginalization, evaluating the marginal likelihood on a grid of intrinsic parameters, and then constructing a fast interpolator to explore the posterior. Marginalization techniques are also used for quickly obtaining subsets of parameters such as for source localization~\cite{Singer:2015ema, Hu:2021nvy, Tsutsui_2021}. 

Machine learning methods based on generative models, such as \texttt{DINGO}~\cite{Dax:2021tsq,Dax:2022pxd, Dax:2024mcn,Gupte:2024jfe} and {other variants~\cite{Polanska:2024zpn}} based on normalizing flows~\cite{kobyzev2020normalizing,papamakarios2021normalizing} and \texttt{Vitamin}~\cite{Gabbard:2019rde} based on variational autoencoders~\cite{pagnoni2018conditionalvariationalautoencoderneural}, have shown great potential in fast PE. Generative models learn the data distribution and can quickly draw samples from the posterior in seconds once trained. The samples can be further improved with importance sampling at an additional time cost of a few minutes~\cite{Dax:2022pxd}. \texttt{DINGO} has been applied to the inference of LVK events and presents fast speed and good accuracy, including the inference with slow waveforms~\cite{Gupte:2024jfe} that could be extremely expensive for traditional Bayesian inference. There are also 3G-specific applications in the literature, such as for overlapping signals~\cite{Langendorff:2022fzq} and long BNS signals in 3G detectors~\cite{Hu:2024oen}. {In addition, methods that can make use of GPU, such as gradient-based samplers~\cite{Bouffanais:2018hoz, Wong:2023lgb} and \texttt{RIFT}~\cite{Wysocki:2019grj}, could also bring significant acceleration compared to traditional samplers.}
The speed advantages of machine learning approaches {and GPU hardware} make them particularly well-suited for catalog-level tasks, a critical need in the 3G era with the huge increase in signal numbers.

\section{\label{sec3}Experiments}
To estimate the cost of Bayesian inference in the 3G era, we need to understand the relationship between signal properties (duration, SNR, etc.) and the associated sampling cost.  To do this, we perform 1200 PE runs with varying settings and fit the resulting relation.
In this section, we describe the experiment setup, outline the process for obtaining the time cost fitting, and discuss the accuracy of different acceleration methods.

\subsection{\label{sec31} Setup}
We consider a detector network contains one triangular 10km-ET operating at the ET-D design sensitivity~\cite{Hild:2010id} at the Virgo site, and one 40km CE detector with the CE2 design sensitivity~\cite{reitze2019_CosmicExplorerContribution} at the LIGO Hanford site. Detector noise is assumed to be stationary and Gaussian, and different events have different noise realizations. The starting frequency is 5\,Hz and data is sampled at 2048\,Hz. 

We simulate three types of binary black hole (BBH) sources: GW151226-like (14+8$M_\odot$, duration=256s), GW150914-like (36+30$M_\odot$, duration=64s), and GW190521-like (85+66 $M_\odot$, duration=16s). They represent typical low, medium, and high mass BBHs in the population and produce long, medium, and short duration signals, respectively. {We fix their intrinsic parameters to the median values given in GWTC-3~\cite{KAGRA:2021vkt} and GWTC-2.1~\cite{LIGOScientific:2021usb} results} and randomly simulate extrinsic parameters except for coalescence time and luminosity distance. The coalescence time is fixed at GPS time $t_c=0\,s$ as this choice should not influence the result. The luminosity distance is scaled to adjust the {optimal} network SNR to values in the set $(12, 30, 100, 500, 1000)$, ranging from near-threshold detections to the high-SNR golden events. For each SNR and source type, 20 events with different extrinsic parameters are simulated. Each event is analyzed with four PE methods: the ``standard'' method (no acceleration method applied), RB, MB, and ROQ, resulting in a total of 1200 PE runs. 

The signals are simulated and analyzed with the \texttt{IMRPhenomPv2}~\cite{Husa:2015iqa,Khan:2015jqa} waveform model, which leads to 15 parameters to estimate. Although \texttt{IMRPhenomPv2} is not the most accurate waveform model to date, it is very fast to evaluate and can be efficiently parallelized for calculation.
{We may hope that by the time of 3G observing, advances in hardware and algorithms will reduce the cost of waveform evaluation, so this can be considered a conservative choice.}
%making it a good representation of the expected waveform evaluation speed in the future {(we expect that more complex waveform models will be fast to calculate in the future thanks to the improvements on hardware and algorithms)}. 
{The prior distributions for the parameters are as follows:}
\begin{itemize}
    \item Chirp mass $\mathcal{M}$: Uniform in $[8.5, 9.5]M_\odot$, $[26, 34]M_\odot$, and $[50, 100]M_\odot$, for  GW151226-like, GW150914-like, and GW190521-like sources, respectively.
    \item Mass ratio $q$: Uniform in $[0.25, 1]$.
    \item Spin dimensionless magnitudes $a_1, a_2$: Uniform in $[0, 0.8]$. 
    \item Spin tilt $\theta_1, \theta_2$: Uniform in cosine (isotropic spin). 
    \item Spin angles $\varphi_{12}, \varphi_{JL}$: Uniform in $[0, 2\pi]$ rad. 
    \item Inclination angle $\theta_{JN}$: Uniform in cosine (isotropic orientation). 
    \item Right ascension $\alpha$: Uniform in $[0, 2\pi]$ rad. 
    \item Declination angle $\delta$: Uniform in sine (isotropic on the sky). 
    \item Polarization angle $\psi$: Uniform in $[0, \pi]$ rad. 
    \item Coalescence phase $\phi_c$: Uniform in $[0, 2\pi]$ rad. 
    \item Coalescence time $t_c$: Uniform in $[-0.1, 0.1]s$. 
    \item Luminosity distance $d_L$: Uniform in the volume between $[d_L/2, 2d_L]$, where $d_L$ is the injected value. 
\end{itemize}

{We use the PE library \texttt{Bilby} v2.2.2~\citep{bilby_zenodo} for our analysis.} We employ the sampler {nessai v0.8.1} for faster convergence, though future improvements to the sampler may further enhance performance. We set the number of live points \texttt{nlive}=2000. Larger \texttt{nlive} may be required for higher SNRs or narrower posteriors such as for BNS signals, which would increase the sampling time. {Experiments are run with \texttt{ncpu}=32 CPU cores with the AMD EPYC 7443 CPU model (base clock 2.85 GHz). The likelihood is evaluated in parallel but the training and sampling parts of \texttt{nessai} are not parallelized and only use one core. As a result, the total sampling CPU core time is calculated as total sampling wall time + (\texttt{ncpu}-1)$\times$likelihood evaluation wall time~\footnote{The sampler \texttt{nessai} provides total sampling wall time (the wall time to finish the whole analysis) and wall time spent evaluating likelihood. Since the likelihood is evaluated in parallel while all other processes use a single CPU core, the total CPU time is calculated as (total sampling wall time-likelihood evaluation wall time)*1 + likelihood evaluation wall time*\texttt{ncpu}, which simplifies to the form in the main text.}. The time cost statistics are directly read from the \texttt{nessai} log files.}

The ROQ bases are calculated using \texttt{PyROQ} package~\cite{Qi:2020lfr}, with the same prior used in PE. Using reconstruction error tolerance of $10^{-8}$ for ROM bases and $10^{-10}$ for quadratic bases, we obtain 392, 183, 136 ROM bases, and 184, 96, 112 quadratic bases for GW151226-like, GW150914-like, GW190521-like sources, respectively. The data length is significantly compressed compared to the original lengths of 64k, 32k, and 16k data points. With a single CPU core, ROQ training takes less than one hour for GW150914-like and GW190521-like sources, but takes five days for the GW151226-like source. {We note that the construction can be parallelized but we do not implement it in this work.} Although it is possible to construct ROQs with a wider prior for more general applications, this would increase the training time and reduce the compression rate. Therefore, we focus the ROQ construction on the simulated signals. The precalculation time cost for building the ROQs is not included in the ROQ sampling time.

The fiducial parameter for RB is chosen to be close to the injection parameter, {roughly 10 times more accurate than the pre-estimates from current matched filtering search~\cite{Villa-Ortega:2022qdo}.}
The error in $\mathcal{M}$ is randomly sampled within $\pm 0.1\%$, while errors for $q$ and $d_L$ are within $\pm 5\%$. The errors for $\alpha$, $\delta$, and $\theta_{JN}$ are sampled within $\pm 10\%$, and the coalescence time has no error ($0\%$). Parameters such as spins and angles, which are difficult to estimate accurately before PE, are set to zero in the fiducial parameters. {All random errors are sampled uniformly within their ranges.}
{Although we choose the error to be 10 times more accurate than current search pipelines can provide~\cite{Villa-Ortega:2022qdo}, they may not be accurate enough for the 3G detections as we observe convergence issues in PE with RB in our experiments, particularly in high-SNR events.} RB is effective only within a region near the fiducial parameters, and it may fail to reach the narrow posterior region when the fiducial parameter is biased. In our experiments, five RB runs failed to converge and were excluded from the time cost estimate. 

%To improve accuracy, we apply importance sampling to reweigh the RB PE results after the direct sampling. The reweighting process takes only less than a minute, and we do not include the reweight time in the RB sampling time.

{With our configuration, all methods produces $\sim 12000$ posterior samples when the sampler converges.} The code for the experiments can be found in the GitHub repository \texttt{3gpemethods}~\footnote{\url{https://github.com/MarinerQ/3gpemethods}}.

\subsection{\label{sec32} Bilinear fitting for time costs}
\begin{figure*}
\includegraphics[width=0.98\textwidth]{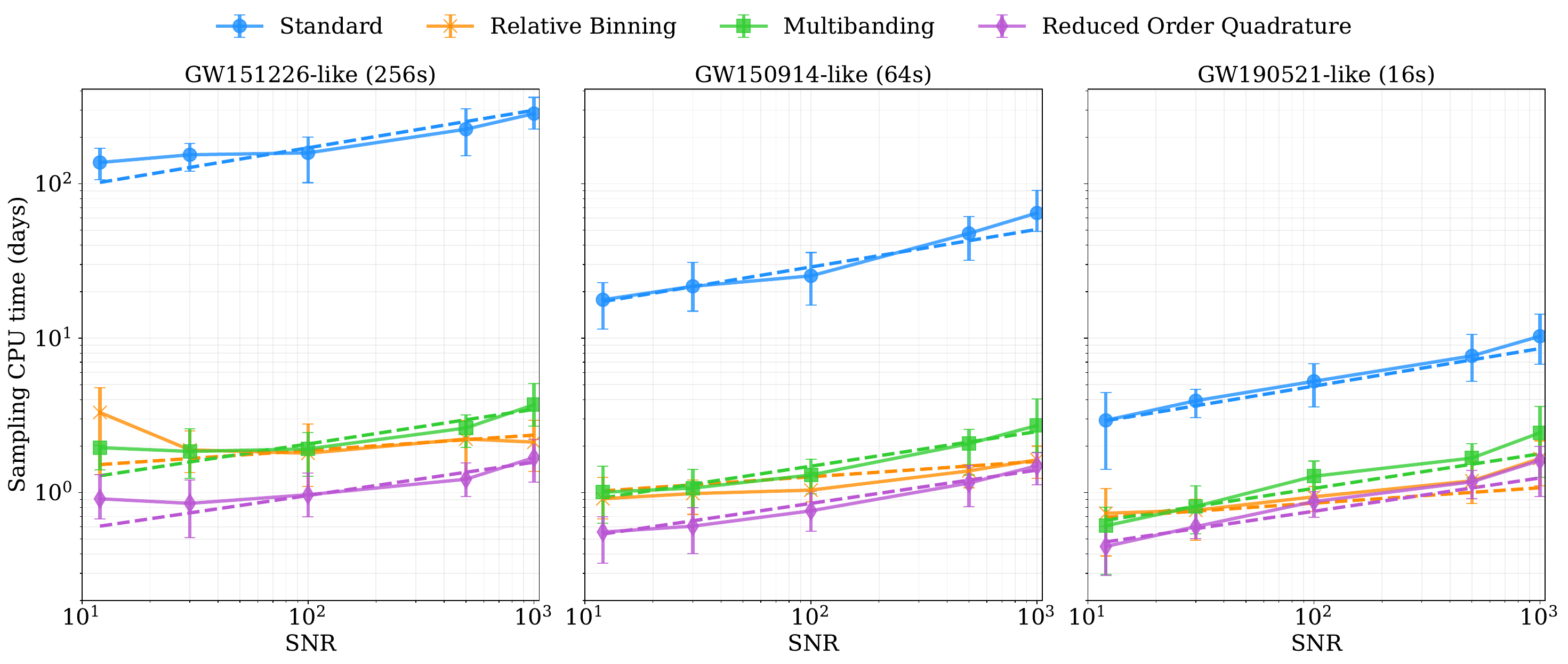}
  \centering
  \caption{\label{fig:time_cost_comp} The CPU core time of stochastic sampling in our experiments,  along with the bilinear fitting. The error bars represent the 16\% to 84\% percentiles of the sampling time across 20 events for each configuration, with the mean indicated in the center. The dashed lines correspond to the bilinear fits of the sampling time cost.}
\end{figure*}

We calculate the CPU core time for stochastic sampling in our experiments, and the results are shown in Fig.~\ref{fig:time_cost_comp}. Overall, ROQ exhibits the fastest speed among all PE methods, completing sampling within 2 CPU days for most events in our simulation. 
MB and RB typically finish within 5 CPU days. In contrast, the standard method can take tens to hundreds of CPU days. Acceleration methods provide a speed-up factor of up to several hundred times, especially in high-SNR and long-signal scenarios.

There is a clear trend that sampling time increases with both SNR and signal duration, with the scaling being approximately linear in the logarithmic scale. Therefore, we employ the following bilinear function to fit the relation between sampling CPU days $D$, signal duration $T$ (in second) and network SNR:
\begin{equation}
    \label{eq:bilinear}
    \log_{10}D = a\log_{10}T + b\log_{10}\mathrm{SNR} + c,
\end{equation}
where the coefficients $a,b,c$ for different PE methods are determined by linear regression {using the \texttt{scikit-learn} package~\cite{scikit-learn}} and are listed in Table~\ref{tab:coeff}. We also calculate the $R^2$ score, the coefficient of determination, a higher value of which indicates a better fit. The $R^2$ score exceeds 0.5 for all PE methods except RB, which may be caused by its convergence issues since the choice of fiducial parameter has impact on the convergence speed. {The relatively low $R^2$ scores in the three acceleration methods implies their run time may depend on signal SNR and duration in a more complex way, but from the regression results in Fig.~\ref{fig:time_cost_comp}, the bilinear fitting is able to capture the overall tendency and the accuracy should be sufficient for a rough, order-of-magnitude level estimation for the PE configuration used in this work. }

%As a crosscheck of the extrapolation, Ref.~\cite{Smith:2021bqc} reported a 90-minute long BNS signal with SNR of 2400 can be analyzed with 1600 CPU core hours using the ROQ technique, while our fitting predicts $\sim 1400$ CPU hours. Given that we are underestimating the time cost (see discussions in Sec.~\ref{sec42}), we consider this a good agreement.

\begin{table}[h]
    \centering
    \begin{tabular}{|c|c|c|c|c|}
    \hline
    Method & $a$ & $b$  & $c$ & $R^2$ \\
    \hline
    Standard & 1.28$\pm$0.02 & 0.24$\pm$0.01 & -1.34$\pm$0.04 & 0.94 \\
    RB & 0.28$\pm$0.02 & 0.10$\pm$0.02 & -0.61$\pm$0.05 & 0.40 \\
    MB & 0.24$\pm$0.02 & 0.22$\pm$0.01 & -0.71$\pm$0.04 & 0.60 \\
    ROQ & 0.09$\pm$0.02 & 0.22$\pm$0.01 & -0.65$\pm$0.04 & 0.51 \\
    \hline
    \end{tabular}
    \caption{\label{tab:coeff}Coefficients $a,b,c$ {and their uncertainties} from bilinear regression in Eq.~\ref{eq:bilinear} for sampling time. The last column gives the coefficient of determination $R^2$ score ranging from 0 to 1, with higher value indicating a better match between prediction and data. }
\end{table}

Looking into the coefficients, we notice the standard method has the largest scaling factors ($a$ and $b$) with respect to the signal duration and SNR, indicating that its time cost grows most rapidly, which aligns with our expectations. ROQ has the lowest scaling factor (the coefficient $a$) with the signal duration, highlighting its good efficiency in speeding up likelihood evaluations.
RB shows the lowest scaling factor (the coefficient $b$) with the signal SNR. 
For the runs that finish quickly (for example, low SNR and short-duration events), we note that the total cost is not dominated by the likelihood but by other operations of the \texttt{nessai} sampler.
%, likely because RB focuses on the parameter space near the fiducial parameter, which is typically close to the true parameter in our experiments. 
%Additionally, RB calculates the ratio between the newly-proposed waveform and fiducial waveform (i.e. heterodyning), canceling the SNR factor. Other methods explore the same likelihood space with the same sampler and only differ on the speed of likelihood calculation, so they have similar $b$ values. 

It is worth mentioning that the effect of lower frequency cutoff $f_\mathrm{low}$ can be estimated roughly using $a$. If $f_\mathrm{low}$ changes to 2Hz or 3Hz, the SNR would not change significantly, but the duration of the signal would scale as $f_\mathrm{low}^{-8/3}$. Using Eq.~\ref{eq:bilinear}, the sampling time $D$ will be \textit{multiplied} by a factor:
\begin{equation}
    \label{eq:flowscale}
    D_{f_\mathrm{low}} = \left(\frac{5}{f_\mathrm{low}}\right)^{\frac{8a}{3}} D_\mathrm{5Hz},
\end{equation}
where $D_\mathrm{5Hz}$ is the result given by Eq.~\ref{eq:bilinear} and $a$ is given in Table \ref{tab:coeff}. Substituting the values, the PE time cost of the standard method would increase roughly 22 and 6 times for $f_\mathrm{low}$ of 2Hz, 3Hz, respectively. For the accelerated methods, the time increase is between 10\% and 100\%. Note that the increase is underestimated, since the changes in SNR are not accounted for. 

\subsection{\label{sec33} Accuracy}
We assess the accuracy of the acceleration methods by comparing the posterior distributions obtained using each method to those obtained with the standard method. For each parameter in each event, we compute the Jensen-Shannon Divergence (JSD) between the posterior distribution from an acceleration method and the standard method. The JSD is calculated with a base of 2, so it ranges from 0 to 1, with the unit of \textit{bits}. A JSD value of 0 indicates that the two posterior distributions are identical, while a value of 1 means that they are completely different, with no overlap in the result. {Ref.~\cite{Romero-Shaw:2020owr} points out that the \texttt{bilby} package produces posterior samples with $\leq 0.0015$ nat ($\approx 0.001$ bit) of statistical fluctuations. Based on this, we consider two sets of posterior samples identical when their JSD is less than $10^{-3}$. For two sets of samples with JSD above this threshold, it is possible to use importance sampling~\cite{Dax:2022pxd} to reweight the approximated posterior samples if they are not significantly biased. Therefore, we consider JSD at the order of $10^{-2}$ ``acceptable'' in this work. }

\begin{figure*}
    \subfigure[GW151226-like]{\includegraphics[width=0.8\textwidth]{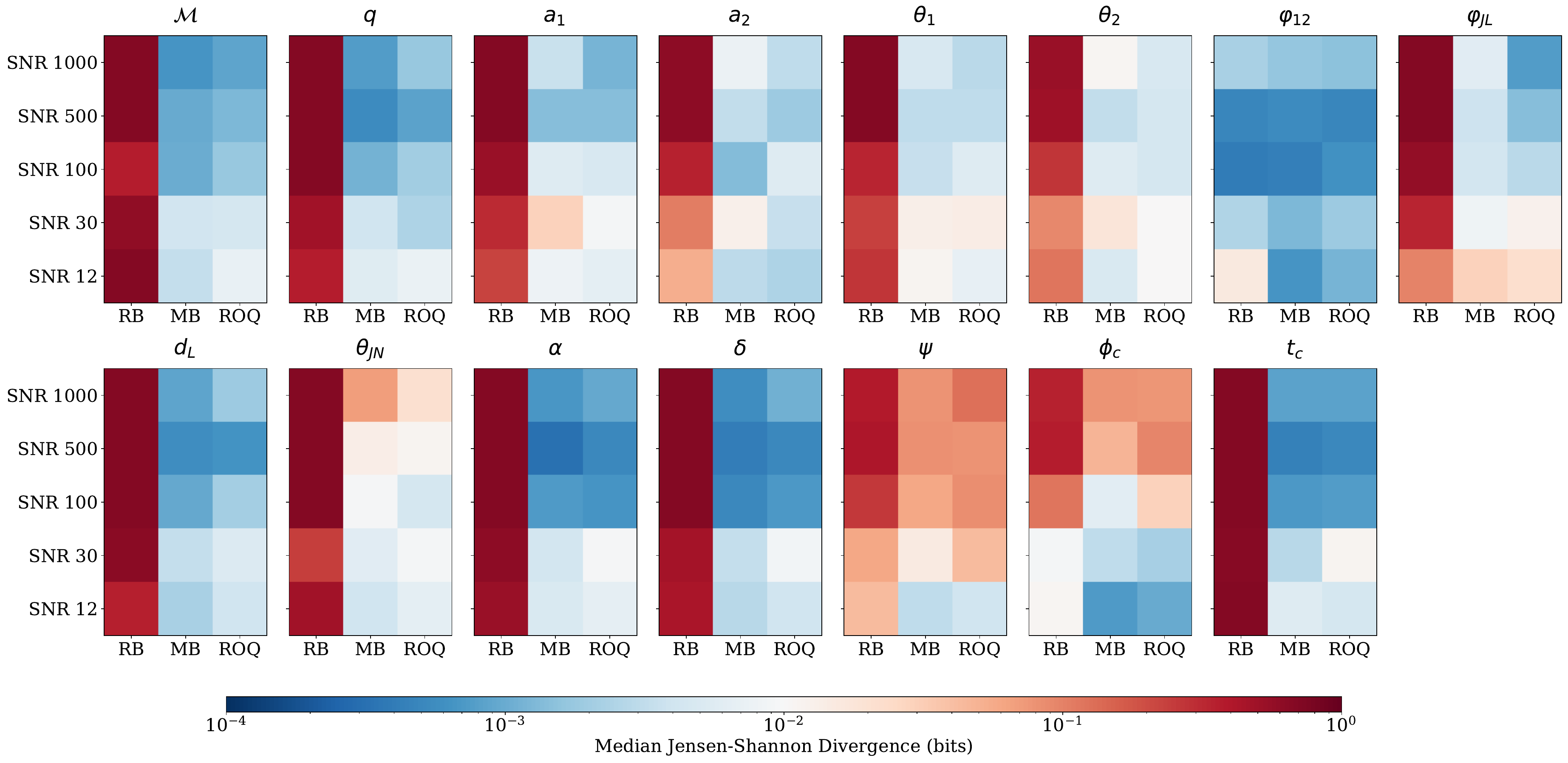}}
    \subfigure[GW150914-like]{\includegraphics[width=0.8\textwidth]{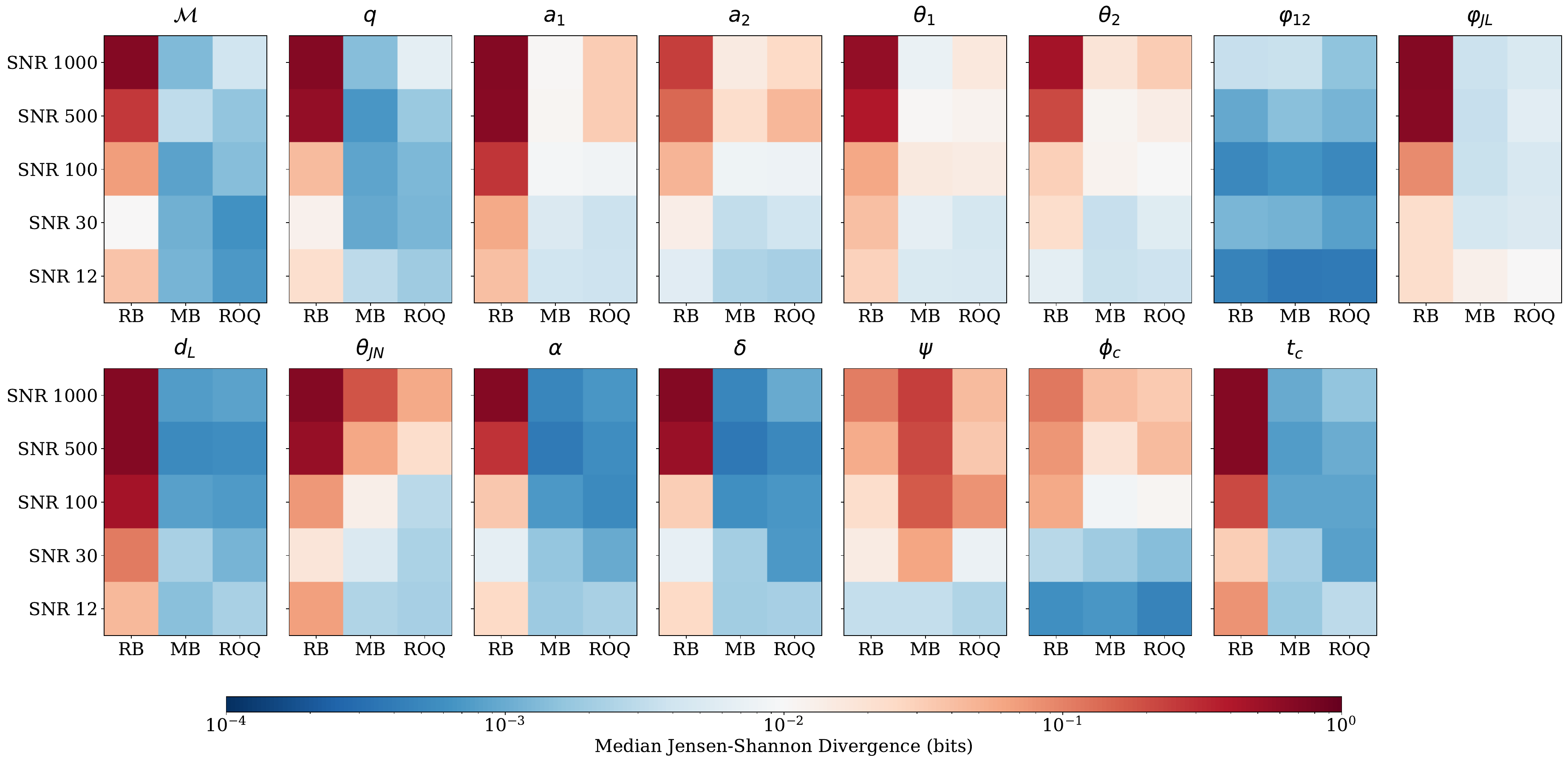}}
    \subfigure[GW190521-like]{\includegraphics[width=0.8\textwidth]{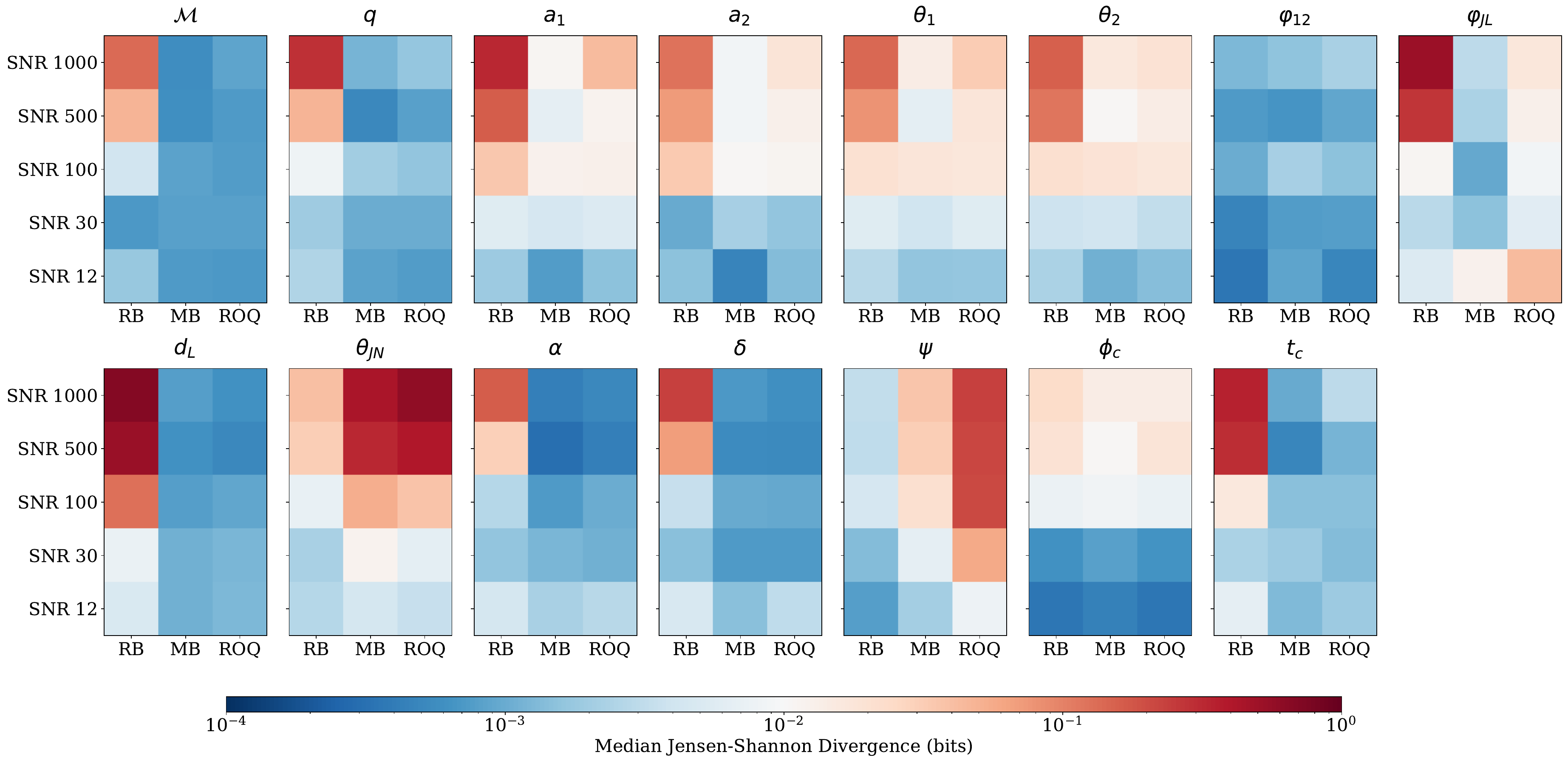}}
    \centering
    \caption{\label{fig:jsd} The Jensen-Shannon Divergence (JSD) between posteriors obtained using acceleration methods and the standard method. Each subfigure corresponds to a different source type, and each patch represents a different parameter. The color of each grid represent the value of JSD in \textit{bits}. A JSD of 0 indicates that the two posterior distributions are identical, while a JSD of 1 indicates they are completely different.}
\end{figure*}

The Jensen-Shannon Divergence (JSD) is shown in Fig.~\ref{fig:jsd}. For ROQ and MB, most parameters are estimated with good accuracy {under SNR of 100, with typical JSD below or slightly above $10^{-3}$ bits. }
%However, the angle between total angular momentum and orbital angular momentum, $\varphi_{JL}$, is often incorrectly estimated. This may stem from errors in the likelihood approximation, as accurate measurement of spin precession requires a more precise likelihood. 
We observe that the error increases with SNR: {all acceleration methods fail to produce identical posterior samples in high-SNR scenarios, especially for spin-related parameters, polarization angle and the coalescence phase. } This means that the likelihood needs to be approximated with a higher accuracy in high SNR cases. For instance, increasing the number of bases in ROQ may help address this issue. RB shows the poorest accuracy among all PE methods, despite having relatively accurate fiducial parameters in our experiments, {suggesting more accurate fiducial parameters are required.} Its accuracy deteriorates when the posterior is narrow, which happens in high SNR events and long signals (where the chirp mass posterior is narrow). Extra work is needed to determine the required accuracy of the fiducial parameter before performing a full PE. A comprehensive investigation of PE accuracy in the 3G era is beyond the scope of this paper and should be addressed in future work. 

\section{\label{sec4}Costs estimates for 3G detectors}
With the estimated time cost given by Eq.~\ref{eq:bilinear}, we now predict the total cost of Bayesian inference for 3G detectors using a realistic source population. We will first introduce the mock data catalog, ET MDC-1, and then demonstrate how the total cost is estimated.

\subsection{\label{sec41} Overview of ET MDC-1}
\begin{figure*}
    \includegraphics[width=0.98\textwidth]{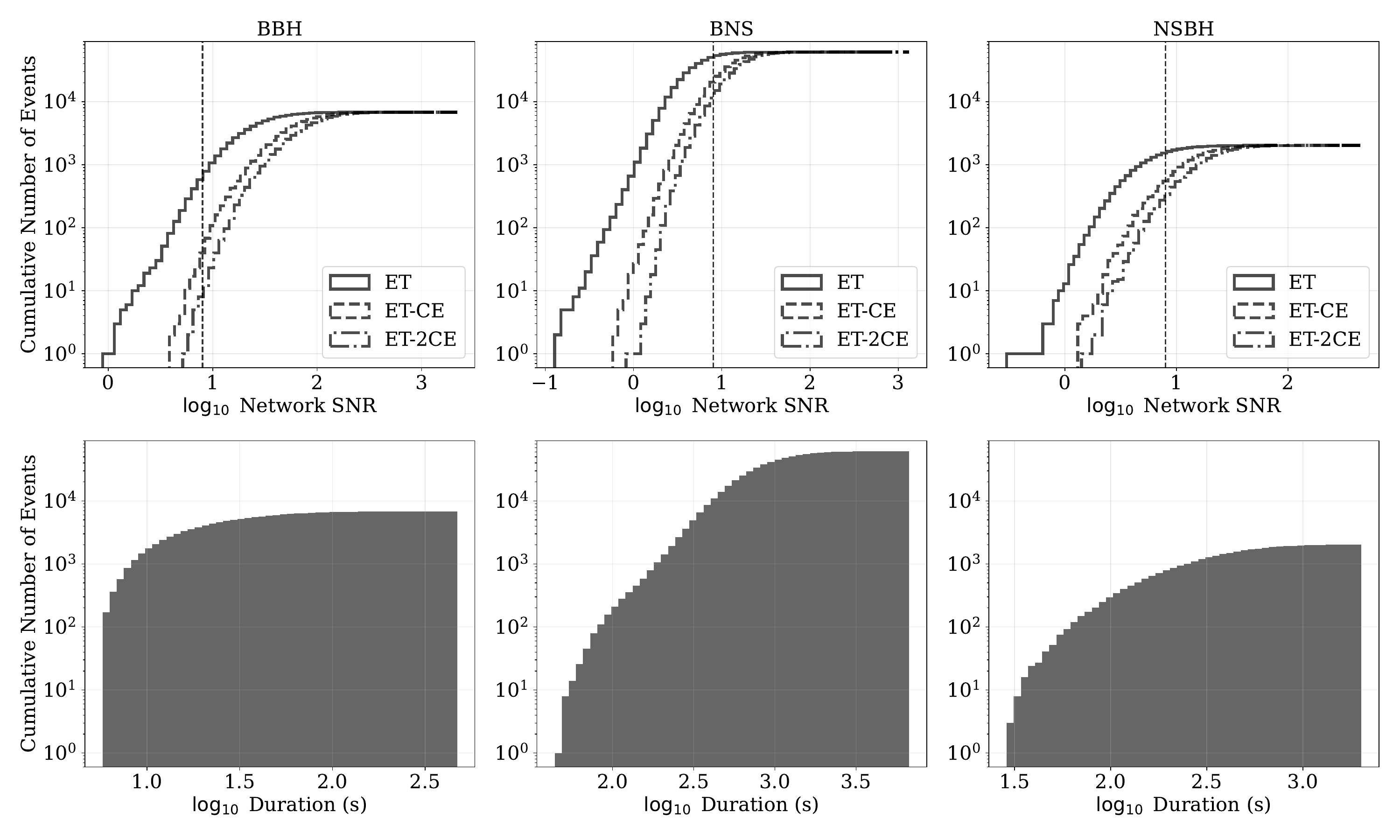}
      \centering
      \caption{\label{fig:mdc_SNR_duration}  \textbf{Top:} Cumulative distributions of network SNR of different sources (in different columns) and different detector networks (in different linestyles). The black dashed lines mark the SNR=8 threshold. \textbf{Bottom:} Cumulative distributions of signal durations, measured from 5 Hz, for different source types.}
\end{figure*}

ET MDC-1 contains one month of simulated observation with the 3G detector ET~\footnote{ET MDC1 data are publicly available from \url{http://et-origin.cism.ucl.ac.be}}. 
The population model used in ET MDC-1 includes 6812 BBH, 61217 BNS, and 2029 NSBH sources over the month, and varying numbers of detections possible depending on the detector network. 
We consider three types of detector networks: ET, ET-CE, and ET-CE-CEL (ET-2CE). Here, `CE' is assumed to be located at the LIGO Hanford site and `CEL' at the LIGO Livingston site, with the same PSD used in the PE experiments. We calculate the network SNR for each network configuration and present the cumulative distribution of the SNR in Fig.~\ref{fig:mdc_SNR_duration}. Setting SNR=8 as the detection threshold, a single ET could detect most BBH sources but would miss a certain fraction of BNS and NSBH sources. Adding CEs to the network significantly enlarges the detection range of BNS and NSBH sources. 

We also show the cumulative distribution of the signal duration in Fig.~\ref{fig:mdc_SNR_duration}. The signal duration is defined as the time it takes for the binary system to evolve from 5 Hz to the merger, computed using the 0PN equation. We add an additional 4s to the duration of each signal to account for the FFT corruption in real data analysis. Most BBH signals have durations less than 100s, while NSBH and BNS signals are mostly longer than 100s. Some BNS signals can last for over an hour. 

Note that ET MDC-1 only provides the GW strain data for ET. However, in this study, we focus solely on the signal duration and the (optimal) SNR, which are independent of the actual noise realization.

\subsection{\label{sec42} Total PE cost estimates}
Instead of directly using GW strain data, we simply use the signal durations and SNRs based on the MDC catalog to predict the sampling time with Eq.~\ref{eq:bilinear}. However, we have overlooked some factors that also affect the sampling time. We discuss their expected contributions here:
\begin{itemize}
    \item Parameter space volume: Our experiments are based on 15-D PE for BBH systems. For NSBH and BNS systems, there are one and two more parameters for the tidal deformability, respectively. {While the astrophysically restricted spin of neutron stars can make PE faster (which, however, is not always used in PE, e.g., Refs.~\cite{abbott2019_PropertiesBinaryNeutron, abbott2020:GW190425ObservationCompact}), the additional tidal parameter could counteract the speedup. } Moreover, future analyses may incorporate eccentricity, adding two more dimensions. {Detector calibration parameters~\cite{LIGOScientific:2016xax} are also ignored in our simulation. } Consequently, we are \textit{underestimating} the total time cost for 3G.  The scaling of sampling time with dimensionality is not entirely certain. Although it is shown that the number of likelihood evaluations scales exponentially with the dimensionality for \texttt{nessai}~\cite{Williams:2021qyt}, the exact scaling depends on the new information (relevant to posterior and prior widths) brought by the new dimensions. Some samplers can achieve polynomial scaling~\cite{Handley_2015}, which may mitigate the increase in time cost due to the increased dimensionality.
    \item {Posterior shape: Adding/removing detectors may reform the shape of the posterior by reducing/increasing multimodalities, which affects the convergence of the sampler. Since the bilinear fitting is obtained using the ET-CE network, posteriors with ET only may have a more complex shape which takes the sampler longer to converge than the prediction. Similarly, the ET-2CE computational cost may be overestimated. }
    \item Sampling rate: We use a sampling rate of 2048\,Hz in all experiments, but for 3G analysis this number could increase, especially for signals with higher modes and low mass systems. We are therefore \textit{underestimating} the time cost. 
    \item Sampling algorithm and configuration: Our investigation a variant of the nested sampling algorithm, which is commonly used in current LVK analyses. Our sampler configuration is fixed in all experiments. However, for longer and louder signals, a larger \texttt{nlive} may be required, which slows down the PE. Since our experiments only extend to T=256\,s, we expect this could happen in signals that last for hours. In this sense, we are \textit{underestimating} the total time cost. We also expect that results would vary with different choices of sampling method, for example MCMC does not require the code to sample the entire prior before converging on the posterior and so may run faster if initial parameter estimates are available~\cite{Wouters:2024oxj}.
    \item Overlapping signals: In realistic 3G detector scenarios, signals from different sources could overlap in the time domain~\cite{himemoto2021_ImpactsOverlappingGravitationalwave,pizzati2022_InferenceOverlappingGravitational,relton2021_ParameterEstimationBias,relton2022_AddressingChallengesDetecting,samajdar2021_BiasesParameterEstimation,Hu:2022bji}, requiring a more sophisticated analysis such as joint PE. We do not consider this effect in the experiments, leading to an \textit{underestimate} of time cost. 
    \item Long signal effects: For signals longer than $\sim$ 10 minutes, several additional effects need to be taken into consideration, such as the Earth's rotation~\cite{Essick:2017wyl} and variations in PSD~\cite{Palomba:2005fp}. These may bring more calculations during PE and thus we are \textit{underestimating} of time cost. 
    \item \textit{Ad hoc} ROQ: We restrict the prior of building ROQ to a small range, which  reduces the required number of ROQ bases and lead to \textit{underestimation} of time cost. However, on the other hand, ROQ could be improved with MB~\cite{Morisaki:2023kuq} and so compress the likelihood better, and this is not included in our experiments. 
    \item Duty cycle and evolving sensitivity: GW detectors may not operate continuously throughout the entire month, and their sensitivity may not reach the design level at the start of the observation period. As a result, we may \textit{overestimate} the total time cost. However, the duty cycle can be easily included by applying a constant factor, and the sensitivity does not influence the validity of Eq.~\ref{eq:bilinear}. Therefore, our methods and results can be readily extended to more realistic observational scenarios.
    \item {Hardware improvements: We employ AMD EPYC 7723 CPU model in this work, whose base clock is 2.85 GHz. Future hardware will be improved. }
\end{itemize}
Although it appears we may have been underestimating the time cost, {we expect
that future developments in sampling techniques in coming years}
%it is difficult to predict future advancements in sampling techniques that 
could further accelerate PE. {For example, the \texttt{dynesty} sampler was tuned further in the analysis of GWTC-4~\cite{LIGOScientific:2025yae}, which improved its performance.} Given these caveats, we can say that with current technology, we provide an \textit{optimistic} estimate of the future PE cost.

\begin{figure*}
    \includegraphics[width=0.98\textwidth]{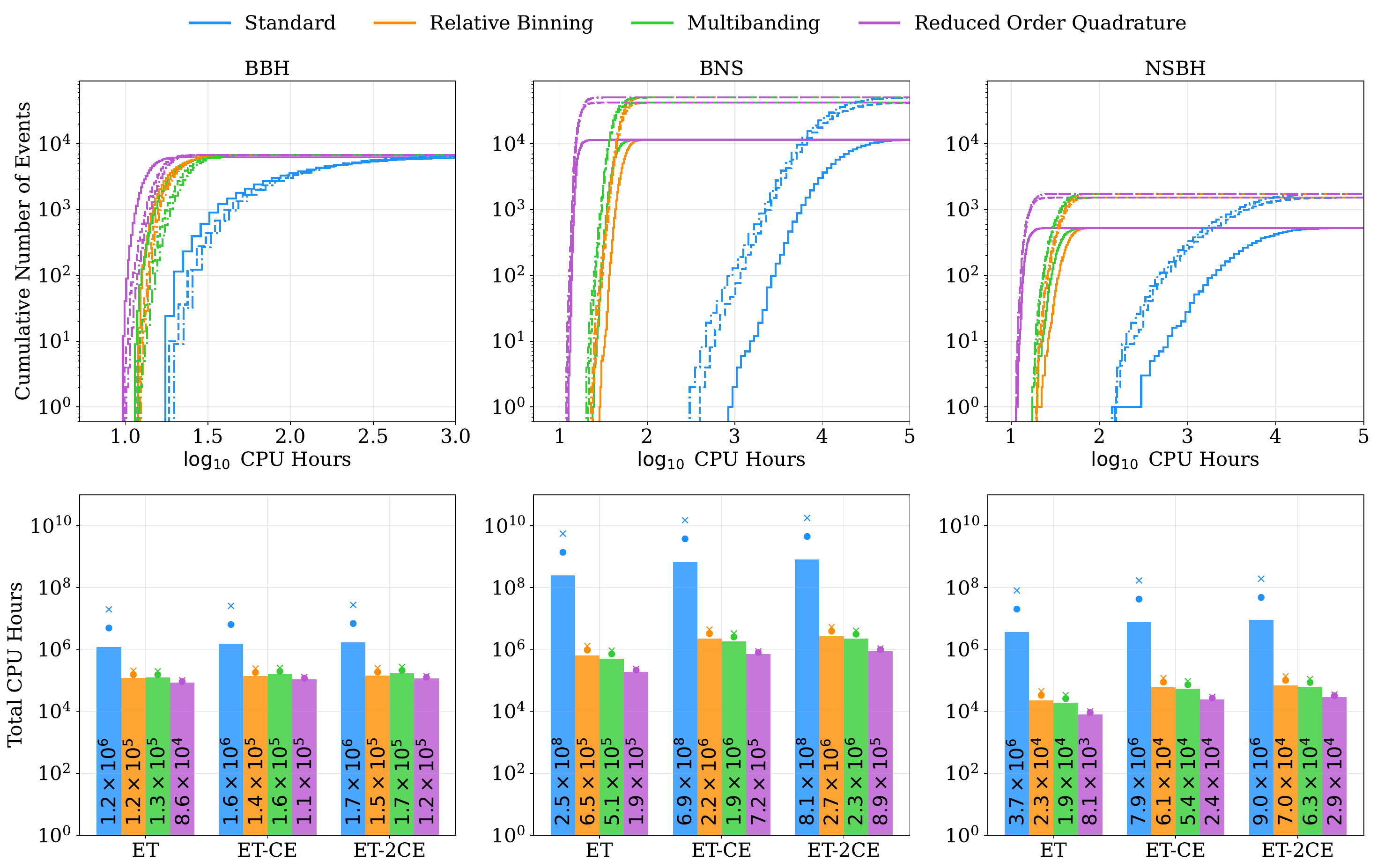}
    \centering
    \caption{\label{fig:mdc_cost} \textbf{Top:} Cumulative distributions of estimated PE cost for detected events (SNR$>$8, in one-month observation) in CPU hours with different detector networks. The solid line represents ET-only network, dashed line represents the ET-CE network, and solid-dashed line represents the ET-2CE network. Different source types are in separate columns, with different colors representing various PE methods. \textbf{Bottom:} Total CPU hours required to perform Bayesian PE for the one-month observation with different detector networks. The numbers indicate the corresponding CPU core hours for each bar. {A dot and a cross above each bar represent the cost for $f_\mathrm{low}=$3Hz and 2Hz, respectively.}}
\end{figure*}

For each source in the MDC that is considered to be detected (network SNR$>$8), we use Eq.~\ref{eq:bilinear} and the effective duration to estimate its PE time cost. The cumulative histogram of the PE time cost is shown in the top row of Fig.~\ref{fig:mdc_cost}. Most events, including BNS and loud signals, can be analyzed within 100 CPU core hours using acceleration methods. %{We note that there could be underestimated by factor of 10 compared with \texttt{pbilby} runs, as discussed in Sec.~\ref{sec42}.} 
However, when using the standard method, the time cost becomes prohibitively large, reaching up to $10^{5}$ CPU core hours. The total time cost to analyze one month of observation is shown in the bottom row of Fig.~\ref{fig:mdc_cost}. Without acceleration methods, the total CPU core hours required can reach $10^{9}$, while acceleration methods can bring this number down to millions. {Further reducing the $f_\mathrm{low}$ from 5Hz to 3Hz or 2Hz, the cost for the standard method scales up drastically, while the acceleration methods do not. Therefore, we should expect the accelerated PE methods to play an important role in the future data analysis. }

We do not fully explore different detector configurations in this study (e.g., L-shaped ET and CE-only networks). However, since the estimation of the PE time cost is purely a function of signal SNR and duration ({i.e., the shape of the posterior is not considered}), we do not expect the total PE time cost prediction to change significantly for different detector configurations. For example, for a single CE detector, the total time cost prediction should lie between the estimations of a single ET and ET-CE network, and a 2CE network should lie between the ET-CE and ET-2CE network. The lower frequency cutoff affects the total PE cost approximately by the factor given in Eq.~\ref{eq:flowscale}, {with examples of 3Hz and 2Hz shown in Fig.~\ref{fig:mdc_cost}. }

The LVK Computing Infrastructure~\cite{bagnasco2023, ligo_ldg} currently operates with fewer than 50 thousand CPU cores. %Since the \texttt{pbilby} is employed in LVK PE production, and Ref.~\cite{Smith:2019ucc} shows that $\sim 10^2-10^4$ CPU hours is required to analyze a event once, 
{Using the PE cost fitting for standard method,} we could infer that at least $10^4-10^6$ CPU hours is required to analyze all (over 200) detections over the past 10 years. It should be noted that this is the cost of performing PE only once; in LVK production, each event may be analyzed multiple times before the final results are obtained for publication. Ref.~\cite{bagnasco2023} shows that all data analysis activities during O3 cost $7.5\times10^9$ CPU hours under the HS06 benchmark~\cite{hs06}, corresponding to $\sim 3\times 10^8$ CPU core hours for the CPU model used in this work. 
Therefore, the millions of CPU hours required for the one-month observation in the future will be a heavy burden for current computing infrastructure. It would take several days to process on the entire computing cluster, consuming approximately 1 GWh of electricity and costing hundreds of thousands of USD in electricity charges, assuming 150W of CPU power and 0.15 dollar per kWh of electricity. 
Considering potential future upgrades to computing clusters (e.g., the Worldwide LHC Computing Grid, which operates with more than 500,000 CPU cores~\cite{HEPSoftwareFoundation:2017ggl}), we conclude that while full Bayesian PE is technically feasible, it is neither budget-friendly nor environmentally sustainable. {Therefore, more efficient PE methods, such as machine-learning-based methods that greatly outperforms traditional stochastic sampling methods discussed in this paper, will be crucial for GW astronomy in the 3G era.}

\section{\label{sec5}Conclusions and discussions}
The computational challenge for the next-generation GW detectors has drawn significant attention in recent years, yet the precise computational cost has been largely an unexplored domain. In this paper, we investigated the computational cost associated with one of the most computationally intensive aspects of GW astronomy: Bayesian PE with stochastic sampling in the 3G era. We started from simulated signals that can be analyzed with a manageable cost and obtained a relationship between the sampling time and signal duration and SNR. With this relationship, we predicted the total PE cost for the event catalog from the ET MDC-1.

Our experiments include the standard PE method and three accelerated methods: relative binning, multibanding, and reduced order quadrature. The standard method has been shown to be impractically slow for the 3G catalog: using this approach, some events would require {$10^{5}$} CPU hours to analyze, and analyzing the entire one-month catalog could demand over billions CPU hours. In contrast, the acceleration methods significantly reduce the time cost. Most events can be analyzed within 100 CPU hours, and the total cost for the entire catalog observed in a month is reduced to millions of CPU hours.

Among the acceleration methods we tested, we found that ROQ provides the best speed while maintaining reasonable accuracy. {However, the choice of ROQ prior may affect the speed~\cite{Morisaki:2020oqk}. While a narrower prior brings faster speed, more sets of ROQ bases would be needed to cover the entire parameter space, which would consume more computing resources to generate in advance.} RB and MB have similar speeds, approximately twice as slow as ROQ. However, we observed accuracy issues with RB, potentially due to its dependency on fiducial parameters. {Assuming not knowing the true parameters, how to assign accurate fiducial parameters needs to be addressed in future work (e.g. whether point estimates derived from detection pipelines are accurate enough.)} We also emphasize that systematic errors in high SNR events need to be carefully controlled. For example, more ROQ bases may be required in high-SNR events for better accuracy. Further investigations into the accuracy of PE in the 3G era are necessary, and we plan to address this in future work.

{As discussed in Sec.~\ref{sec42}, the numbers presented in this work represent an optimistic estimate of extending our analysis configurations to future data. 
However, the potentially underestimated computational cost for the future - millions of CPU hours per month - is sufficient to analyze all LVK detections over the past 10 years. It remains a substantial burden on computing infrastructure, electricity costs, and environmental impact. To address the computational burden in the 3G era, more efficient methods must be employed, such as advanced samplers, improved data compression, faster waveform evaluations, parameter marginalization, and machine learning techniques. }

We hope that the methods and results in this work can serve as a reference and baseline benchmark for the research and development of 3G detectors. 
When novel data analysis methods are developed in the future, whether for more realistic data (e.g. including overlapped signals and noise variations) or a faster speed, their computational cost can be assessed and compared with the results presented here. For analyses at the catalog level, the computing efficiency must be carefully considered, as the cost for processing a single event will be multiplied across the entire 3G catalog.

\begin{acknowledgments}
We would like to thank Tania Regimbau for helping with the use of ET MDC data and thank Michael Williams for helpful discussions about \texttt{nessai}. We also thank Filippo Santoliquido for helpful discussions and Steve Fairhurst for feedback. We are grateful for computational resources provided by Cardiff University, and funded by STFC grant ST/I006285/1. This work was supported by STFC grant ST/Y004256/1. QH was also supported by CSC.
\end{acknowledgments}

\bibliography{refs.bib}

\end{document}